# Screening articles by citation reputation

Vicente Safón [1a, b]

Domingo Docampo [2]

Lawrence Cram [3a, b*]

[1a] Department of Management, University of Valencia, Valencia, Spain

[1b] Valencian Institute of Economic Research, Valencia, Spain

[2] Atlantic Research Center for Information and Communication Technologies, University of Vigo, Vigo, Spain

[3a] Research School of Physics, Australian National University, Canberra, 2600, ACT, Australia

[3b] Charles Darwin University, Brinkin, 0810, NT, Australia

[*] Corresponding author. Lawrence.Cram@anu.edu.au

ORCID:

Vicente Safón            https://orcid.org/0000-0002-1144-5924

Domingo Docampo      https://orcid.org/0000-0001-6864-1232

Lawrence Cram          https://orcid.org/0000-0002-8096-447X

ABSTRACT

We introduce reputable citations (RC), a method to screen and segment a collection of papers by decoupling popularity and influence. We demonstrate RC using recent works published in a large set of mathematics journals from Clarivate's Incites Essential Science Indicators, leveraging Clarivate's Web of Science for citation reports and assigning prestige values to institutions based on well-known international rankings. We compare researchers drawn from two samples: highly cited researchers (HC) and mathematicians whose influence is acknowledged by peers (Control).

RC scores distinguish the influence of researchers beyond citations, revealing highly cited mathematical work of modest influence. The control group, comprising peer-acknowledged researchers, dominates the top tier of RC scores despite having fewer total citations than the HC group. Influence, as recognized by peers, does not always correlate with high citation counts, and RC scores offer a nuanced distinction between the two. With development, RC scores could automate screening of citations to identify exceptional and influential research, while addressing manipulative practices. The first application of RC reveals mathematics works that may be cited for reasons unrelated to genuine research advancements, suggesting a need for continued development of this method to mitigate such trends.

1   INTRODUCTION

A reference listed in a research paper may indicate that the paper's authors wish to acknowledge the usefulness of ideas reported in that reference, establishing a link between the number of citations to a work, and the extent of its influence or value (Garfield, 1979; Leydesdorff, 1998; Tahamtan & Bornmann, 2022, sect. 3.6; Van Raan, 2005; Ziman, 2001, p. 261). Well-recognized influential researchers often have relatively high citation counts and rates per paper (Garfield & Welljams-Dorof, 1992). Moreover, citations are so numerous that even if many do not indicate specific occasions of influence or value, a large number will (Cronin, 2005). However, it does not follow that highly influential authors have the highest citation counts, nor that highly cited authors have the most influence. There are many reasons why papers and authors accumulate citations, most having little to do with intrinsic usefulness (reviewed by Bornmann & Daniel, 2008; Tahamtan & Bornmann, 2019). The seduction of misapplying citation counts as proxies for value has partially inspired research evaluation risk mitigation protocols (Hicks et al., 2015; Pendlebury, 2008; Wilsdon et al., 2015).

A turning point in the use of citation counts to measure research influence occurred in 2001 when the Institute for Scientific Information (ISI) introduced Highly Cited Researchers (HCRs, Thomson-ISI, 2001). The documentation claimed that the highest citation counts in a field identify the world's most influential scientific authors in that field. While the founder of ISI, Eugene Garfield, often used 'citation counts' and 'influence' as synonyms (Garfield & Welljams-Dorof, 1992), Thomson-ISI meant that citation counts are the measure of influence, not a synonym for it. Since then, Thomson Reuters ISI and the subsequent owner Clarivate have been publishing annual lists of HCRs by discipline, while tempering the claim that the list identifies the most influential researchers. For example, the Wayback Machine reveals the *highlycited.com* site shifting the claim from "the most influential authors" in 2001 to "a strong indicator of scientific contribution" by 2013. With the radical redefinition of the HCR category in 2014, ISI further weakened the claimed connection, advising that "the only reasonable approach to interpreting a list of top researchers such as ours is to fully understand the method behind the data and results, and why the method was used. With that knowledge, in the end, the results may be judged by users as relevant or irrelevant to their needs or interests."

More than two decades after the HCR category was launched, Clarivate (2023) excluded the mathematics category from its 2023 HCR list, explaining that "the field of Mathematics is more vulnerable to strategies to optimize status and rewards through publication and citation manipulation". Catanzaro (2024) reported that cliques or citation cartels (see for example Fister et al., 2016) appear to be churning out and citing low-quality papers to improve some authors' and universities' rankings, and suggested that Clarivate may have evidence of citation manipulation in mathematics. When the direct descendent of the ground-breaking Institute for Scientific Information (Garfield, 1955, 1998) decides that the manipulation of citation counts is obscuring influence in a major field such as mathematics, it is timely to search for patterns of citation manipulation in mathematics, and ask why they arise.

Page | 2

The possibility that citations are manipulated to garner undeserved scholarly standing for individuals and journals has been acknowledged since the early days of citation indexing, often alongside the claim that an antidote is the opprobrium that discovery would mean (e.g., Biagioli & Lippman, 2020; Garfield, 1979; Rousseau et al., 2018, pp 106-107). Here, we are particularly interested in manipulations that reflect strategic interactions between citing and cited mathematicians. Interpretivist and functionalist accounts of these interactions are sometimes regarded as oppositional, although both generally regard most interactions as the one-way selection of the cited by the citer (Borgman & Furner, 2002; Ziman, 2001 in a silimiar way, describes the differences in terms of normative and social constructionist viewpoints). An interpretivist lens sees the citer's references as an efficient way to review current knowledge, establish novelty, and demonstrate that the citers' work has addressed a recognized problem. Citers refer to work that is important and correct, and occasionally to work needing correction and they avoid citing trivial or irrelevant work since that could reduce potential readers' respect for their new work (Gilbert, 1977). A functionalist lens sees the citer selecting references to acknowledge priority, novelty, or utility, or occasionally to criticize. The citer is mindful of the social norms of science: to choose objectively and disinterestedly, and not on personal or social attributes of the cited; to recognize and respect the intellectual property rights of earlier authors; and to evaluate their choices with detached care and skepticism (Merton, 1973; pp. 270-278).

Tahamtan and Bornmann (2022) have integrated these and other citation theories in the Social Systems Citation Theory (SSCT). SSCT proposes that citations and publications form a self-organizing social system of science bonded by communication and organized to produce scientific truth. Researchers and other humans are not part of this system, but rather belong to a self-organizing psychic system of ideas and motives. The theory encourages us to separate citation bibliometrics (in the social system of science) and the inference of motives (in the psychic system of researchers) or drivers (in the policy system of institutions). While the systems interact (e.g., ideas are published, publications influence ideas, institutions reward publications) SSCT proposes that they can be analyzed and understood as relatively autonomous systems with linkages. Leydesdorff (1998, p. 9) earlier advanced a similar perspective: "citations are the result of the interaction between networks of authors and between networks of their communications." In adopting this perspective, we first design and implement a bibliometric screen, then explore bibliometric correlates of highly cited works of modest utility, and finally suggest some of the drivers of manipulation from the psychic and policy systems.

A dominant feature of the social system of science is unending growth in the annual rate of production of papers, together with burgeoning linkages in the citation network. The acceleration of paper production arises from the combination of the endless frontier of discovery in the psychic system (Bush, 1945) and the global expansion and massification of higher education in the policy system (Salmi, 2013). Sixty years ago Price (1963 p.6) emphasized the rapidity of growth in article numbers, suggesting a doubling time as short as 10 years. The past 25 years of Scimago data confirm that the mathematics category has an article doubling time remarkably close to 10 years, so the annual production of mathematics papers is now around sixty times greater than it was when Price wrote.

As Price foresaw, growing publication numbers lead to structural change in the publication industry, with more reviews, more journals, and changing aggregation roles (Aspesi et al., 2019;

Fyfe et al., 2017). The policies of a very few commercial publishers, including Elsevier, Springer, Taylor & Francis, Wiley, and Sage now apply to more than half of all scientific publications. Mega journals, such as PLOS One, Scientific Reports or Heliyon have emerged, publishing thousands of papers yearly. New publishers, often with lightweight peer review procedures, strive for the revenue benefit that follows from high rankings in systems such as JCR or Scopus. Paper mills, and predatory or hijacked journals which charge fees for publishing without thorough peer review or simply as scams, are expanding (Abalkina, 2023; Beall, 2012; Kapovich, 2011).

Price (1963) also conjectured that the doubling rate for remarkably high quality papers might be 20 years rather than 10 years. If so, the proportion of very high-quality work published each year would have declined to a few percent of its value in 1963. If such a decline has occurred, it would be hard to discover by traditional citation counting. The distribution of reference counts over a population of works is stable due to page-length limits and these references become widely allocated as citations, masking any disproportionate growth in works of lower impact. To explore the potential de-coupling of citation counts and influence, our paper operationalizes a concept of reputable citations that can screen and segment a collection of papers by influence, even when they all have many citations.

## 1.1 Reputable citations

As noted above, authors have diverse motivations for mentioning a source, and references do not all have the same value as indicators of influence. One approach to constructing a value scale for the influence of references is to examine each in context. For example, a reference appearing only once in the introduction can be assigned a lower value than one mentioned several times in the methods and discussion sections due to its stronger influence, allowing a nuanced view of the value of the referenced work (Small, 1980; Tahamtan & Bornmann, 2019; Wan & Liu, 2014; Zhu et al., 2015). This procedure helps to explain the influence of the referenced work on the citing author but carries no information about the standing of the citing work.

To explore the standing of citing works Bollen et al. (2006) noted that 'status' in a general social context comprises two factors, the number of endorsements ('popularity') and the prestige of the endorsers ('prestige'). They pointed out that counting citations to establish scholarly journal status is about popularity and not prestige. They suggested that the PageRank algorithm (Brin & Page, 1998; Page et al., 1998), which combines popularity and prestige (of the citing journals), could start to change researchers' perceptions of journal and article status. Their comparison of journal status derived from simple counting and from PageRank illustrated the feasibility of the approach (see also Pinski & Narin, 1976). The extensive body of analytical studies of PageRank (e.g., Berkhin, 2005) can be readily applied to the similar problem of establishing article, author, journal or institution ranking from the citation network (Massucci & Docampo, 2019). Kanellos et al. (2019) assess more than thirty studies of article-article citation networks using the PageRank and similar algorithms, in which citations are modified by factors to represent time-delay and/or article metadata including author, journal and institution. The studies aim to improve measures of impact at the article level. While some could be adapted to screen for collective manipulative activity, it would be challenging to use article-level data alone to separate manipulation from the collective activity of the "legitimate invisible college" (Welljams-Dorof, 1997, p. 209).

Page | 4

Any algorithm that uses only link counts (whether between web pages or research journals) to infer popularity and prestige can be influenced by artificial links designed to alter the PageRank scoring distribution. Page et al. (1998) for example refer to PageRank manipulation and abuse by commercial interests. They introduce a parameter E, a controllable external source of rank (status) over web pages, that can be adjusted to maintain the utility of PageRank even when manipulation is occurring. The overlay of additional prior information on the link adjacency matrix has been a major research direction for PageRank-type algorithms to help control manipulation and to improve search outcomes (e.g., Davison, 2000; Gyöngyi et al., 2006; Haveliwala, 2002). Fortunately, as Gyöngyi et al. point out, while accurate prior knowledge about site quality is not usually available, approximations yield useful results. The EigenTrust algorithm (Kamvar et al., 2003) addresses the similar problem of establishing the trustworthiness T of contributing sites in a peer-to-peer sharing network.

The counterpart in citation analysis to knowing E for PageRank or T for EigenTrust is to have prior knowledge about the relative status of journals in the citation network. This knowledge cannot come from analysis of the citation network alone since citation-based status is susceptible to and reflective of any manipulations. Our approach is to inject prior knowledge about the status of journals using the relative prestige of institutions that host publishing mathematicians. This information, which does not need to be complete or precise, propagates iteratively into the citation network ranking via the density of prestigious institutions hosting the papers published in a particular journal. The outcome is a reputation weight for each journal, hence a reputation weight for each reference from that journal (i.e. a reputation weight for each citation in the citation network) and finally an average reputation weight for an author or paper. When this weight is high, we say that the citations are reputable. Reputation serves as a guardian of quality, and those who want to achieve or maintain it, be they authors, journals, or universities, ensure that their products (papers), processes, and inputs (scholars) are of high quality (Docampo and Safón, 2021). Once we have determined reputation weights for citations, the citations to an author can be screened and we can begin to explore those highly cited authors publishing papers with lower reputation rank.

The idea of introducing *exogenous* reputational information has been used previously in the literature on journal rankings. For example, Jarvis and Coleman (1997, 2007) ranked law journals according to the national prominence of the authors of their lead articles. Gorman and Kanet (2005) ranked operations management–related journals using the Author Affiliation Index based on the percentage of U.S. academic authors from top U.S. business research universities publishing in the journal. Chan et al. (2013) ranked finance journals using an Author Concentration Index that uses the percentage of the journal's articles authored by a certain number of leading authors. Docampo and Safón (2021) proposed a classification of journals for 14 social science disciplines based on the prestige of the universities of the authors of the papers published in the analyzed journals. Chen and Chen (2011) combined two reputational elements using an internal source (citation network) and an external source (expert judgment).

Other research has injected into journal rating some additional *endogenous* reputational information from citations themselves. Chughtai et al. (2018) uses the H-index as a weighting factor to reflect author prestige. Zhang (2017) leveraged the H-index and specific patterns found the article-article citation network to quantify journal impact. Safón and Docampo (2023) elaborated a journal ranking in the LIS discipline using institutional Category Normalized Citation

Page | 5

Impact (CNCI). The reputation captured by CNCI scores are passed from institutions to authors, from authors to papers, and from papers to journals, allowing for the creation of a ranking.

So far as we know, no study has fully formalized the idea of weighting citations directly according to the reputation of the citing work (via institution and author) obtained exogenously rather than endogenously from the citation network. However, our idea does echo various research lines in bibliometrics and scientometrics and has been indirectly discussed and explored through previous approaches.

The remainder of the article is organized as follows. In Section 2 we cover the data and methods used to compute reputable citations. In Section 3 we present the results of our analysis of the field of mathematics defined by the journal set used by Essential Science Indicators. Section 4 is devoted to a discussion of the correlates for low reputation ranking among some highly cited authors, while Section 5 closes the paper by reviewing the limitations of our approach and suggesting directions for further work.

## 2. METHODOLOGY

We illustrate the reputable citation (RC) method using a set of 960 journals classified under Mathematics [44% of journals, some in multiple categories], Mathematics (Applied) [29%], Mathematics (Interdiciplinary Applications) [12%], and Statistics & Probability [15%] by Clarivate's Incites Essential Science Indicators (ESI). For every work (article, review, proceedings paper, or book chapter) published in these journals in 2020, we used Clarivate's Web of Science to capture a citation report listing the indexed works in 2020 or later that cite this work. The data set comprises the time-constrained journal set $J$, work set $W$, author set $A$, and institution set $I$ that form a matrix or network coupled through citations. Not all citations to or references from $W$ are conferred on members of $W$ and we will later describe ways to deal with these 'dangling' citations.

We allocated prestige values to the institutions in $I$ by combining seven international institutional rankings in Mathematics or Statistics that do not rely directly on citation-based rankings of mathematics journals or works. We downloaded institutional rankings for mathematics from the official web pages of EduRank (100 institutions in 2024), NTU (494 institutions in 2023), QS (553 in 2024), Research.com (698 in 2023), SCIMAGO (3110 in 2024), THE (374 in 2024), and US News (400 in 2022-23) and from them derived lists of academic, research, governmental, and corporate institutions cross-matched to the Incites institutional names in $I$. For each of the seven sources we rank institutions into deciles with values 1 (top), 0.9, 0.8, .. . Assigning a minimum score of 1.0 to all institutions, we compute an institution prestige score by adding the decile values for all sources. The highest institution prestige score in $p^I$ is 8.0 (2% of all institutions) and the lowest is 1.0 (38%).

- 2.1 Reputable citation algorithm

Prestige rank scores $p^J$ of the 960 journals are scaled to lie between predetermined limits of (min, max) = (0.05, 10.0) to align with the ranges encountered in InCites' Category Normalized Citation Impact (CNCI: 0.05 – 9.47), Journal Citation Impact Factor (JCI: 0.06 – 11.36), and Article Influence Score (ARTF: 0.05 – 10.63). The numerical limits establish the absolute values of weightings for reputable citations.

The reputable citation algorithm computes journal prestige scores $p^{J(k)}$ recursively over $k$ with unity as the initial values $p^{J(k=0)} = 1$. The recursion proceeds as follows:

At step $k$ for all journals $j$ in $\mathbf{J}$:

- compute the weight $w_{j,i,l}^{(k)}$ for a citation to a work published in journal $j$ (cited journal) from a work hosted by institution $i$ of $\mathbf{I}$ and published in journal $l$ (citing journal) as the product of the fixed institutional prestige value $p_i^I$ and a journal prestige value $p_l^{J(k)}$ updated by recursion:
$$w_{j,i,l}^{(k)} = p_i^I p_l^{J(k)} \qquad (1)$$

- update the relative prestige $\tilde{p}_j^{J(k+1)}$ of journal $j$ conferred by all citations to its works from the set of citing journals to papers published in journal $j$ in 2020, $L_j$. The summation covers all citations, including those between different works in the same journal and those from works published outside the journal set, and is normalized by the number of papers published by journal $j$ in 2020, $N_j$:
$$\hat{p}_j^{J(k+1)} = \frac{1}{N_j} \sum_{l=1}^{L_j} w_{j,i,l}^{(k)} f_i \qquad (2)$$

- linearly rescale $\hat{p}_j^{J(k+1)}$ to lie between 0.05 and 10.0,
$$p_j^{J(k+1)} = 0.05 + 9.95 \left( \frac{\hat{p}_j^{J(k+1)} - \max_J(\hat{p}_j^{J(k+1)})}{\max_J(\hat{p}_j^{J(k+1)}) - \max_J(\hat{p}_j^{J(k+1)})} \right) \qquad (3)$$

The recurrence continues ($k \to \infty$) until the difference between the mean journal score at two consecutive iterations falls below a predetermined threshold. Around thirty iterations reach a difference below 0.0001.

The factor $f_j$ is used to explore citations from works not published in the journal set, and thus not subject to adjustment due to institutional prestige in mathematics. Highly cited researchers influence others no matter their field. Thus, there are highly cited mathematicians who receive most citations from other fields while others are cited for work driving the core of the field of mathematics itself.

When the journal $j$ is in the journal set, $f_j = 1$ and the citation always counts fully. When a citing article appears in a journal $j$ that is not in the journal set, we use $f_j = 1$ or $f_j = 0$ to denote, respectively, full counting or omission of its citations. The scenarios offer complementary views of influence, although the second provides the direct, relevant path to measure in-field influence.

To explore the construct validity of the RC procedure we compare in Figure 1 the derived journal RC score with InCites' 5-year Journal Impact Factor (JIF5), Journal Citation Indicator (JCI) and Article Influence Score (AIS). JIF5 and JCI both use unweighted citation counts, JCI being field-normalized, while AIS, like RC, takes account of the influence of the citing journal. For this comparison, like AIS the RC score includes citations from exogenous journals and does not include institutional prestige. The Spearman rank coefficient between AIS and RC is 79% indicating that these indicators are similar measures of relative influence in the journal set.

**Figure 1**. 5-year Journal Impact Factor (JIF5), Journal Citation Indicator (JCI), and Article Influence Score (AIS) versus Reputable Citation (RC) score without citations from exogenous journals and without differential institutional prestige.

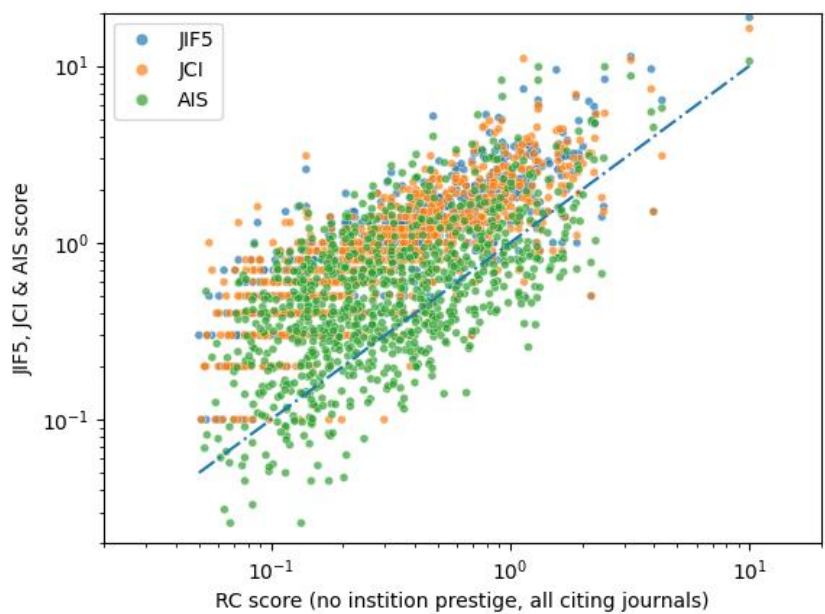

Figure 2 exhibits the journal-by-journal response of RC scores to the exclusion of institutional prestige and the exclusion of citations from exogenous journals, compared with the reference RC model that *includes* institutional prestige and *excludes* citations from exogenous journals. When institutional prestige is excluded, as expected the scores of low-scoring journals increase, while scores of high scoring journals decrease. When citations from exogenous journals are included, the scores of many journals across the range of RC scores are relatively unaffected. However, the scores of some low-scoring journals increase markedly since they receive many additional (unweighted) citations: journals with many articles in applied statistics or mathematical biology are exemplars.

Figure 2. Log-log plot of RC scores the consequences of excluding institutional prestige (yellow), including exogenous citations (green), or both (blue).

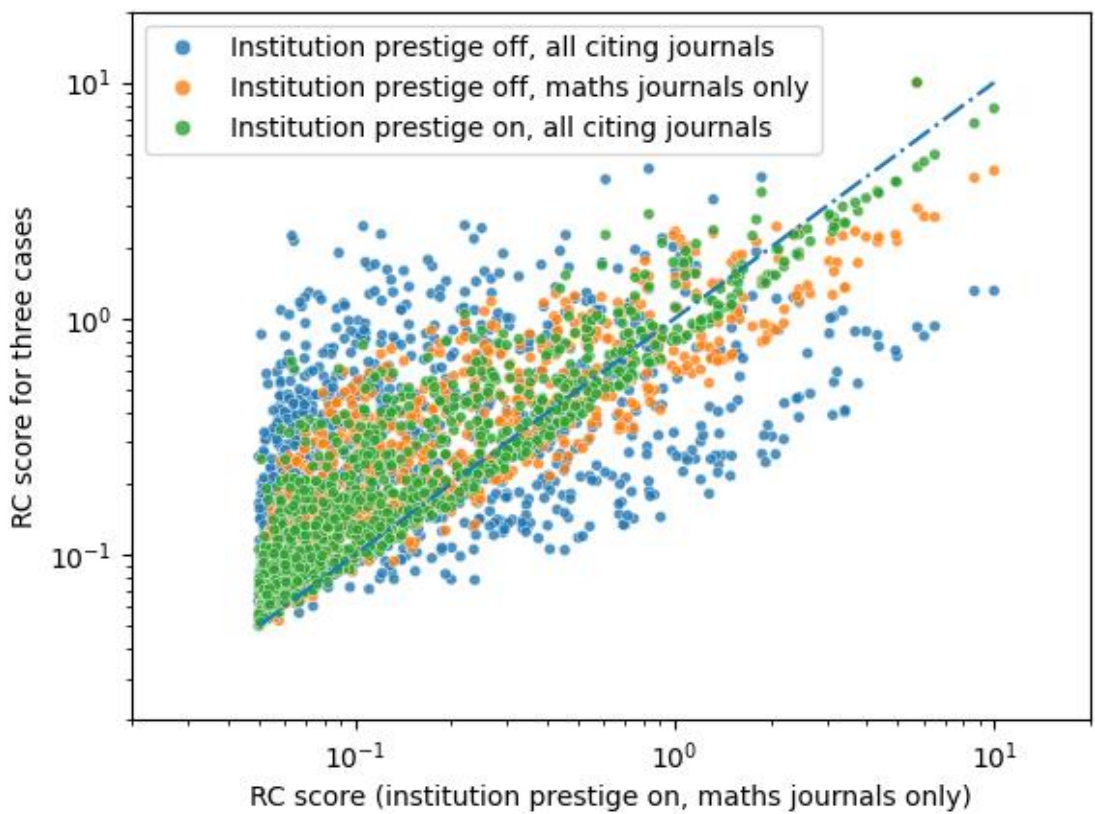

## 2.2  Individual research influence

To compute the reputable citation influence $RC_a$ of an author $a$ who has published $R_a$ mathematics works in the target period and received $L_a$ citations to them, we combine the converged journal prestige $p_j^{J(k \to \infty)}$ and institutional prestige of each article $l$ that cites the author to compute the article-specific converged weight $w_l^{(\infty)}$. Then

$$RC_a^A = \frac{1}{R_a} \sum_{l \in L_a} w_l^{(\infty)} f_l \qquad (4)$$

We may say that the RC score of an author is highly attenuated (weights $w$ are systematically small) when their mean citation counts per paper is larger than their RC score: $L_\alpha / R_\alpha > RC_a^A$.

To establish the viability of the reputable citation algorithm for individual authors, we selected a sample of 300 researchers comprising a Highly Cited and Control group of 150 each:

Highly Cited (HC): researchers with at least ten ESI Highly Cited Papers in Mathematics between 2012 and 2022. In principle these authors would cover the field of promising candidates for nomination as Clarivate HCRs in Mathematics, since somewhat fewer than 100 have been nominated in recent years.

Control: peer-acknowledged researchers awarded a prestigious Mathematics prize (e.g., Fields Medal, Abel Prize) or a European Research Council grant at the ERC Starting, Consolidator, or Advanced levels, topped up to 150 total members with mathematicians who we know are Fellows of a Learned Academy.

Only three Control researchers have ten or more highly cited papers in the fiducial time interval. Since the Controls sample mathematicians whose outstanding contribution is acknowledged by peer review, ideally the RC algorithm will not differentially attenuate the standing of their raw citation scores. On the other hand, the HC group may reveal a range of attenuations if there is a spread of status or standing of their works. To assess this proposition, we sort the RC scores of the two samples and count the number of Control and HC mathematicians in three equal tiers as shown in Table 1.

Table 1. Prevalence of Control and Highly Cited (HC) mathematicians in three equal bins from the list sorted by RC score.

| Sample  | Tier 1 (1-100) | Tier 2 (101-200) | Tier 3 (201-300) |
|---------|----------------|------------------|------------------|
| Control | 93             | 57               | 0                |
| HC      | 7              | 43               | 100              |

If influence were spread evenly across RC scores, each bin in Table 1 would have equal numbers of Control and HC members. However, the Control group dominates the first tier, despite the presence of very few highly cited researchers in the group. On the other hand, despite their extremely high total citation counts the HC group dominates the lowest tier. Their highly cited status differentially excludes them from the first tier. We infer that a comparison of raw citation counts and RC scores delineates the degree of influence of mathematicians as acknowledged by peers. Strongly attenuated RC scores for highly cited mathematicians identify a corpus of work that is less influential than its raw citation counts might suggest. On the other hand, the rarer case of weakly attenuated RC scores for highly cited researchers suggests a body of work that is highly influential and highly cited.

3. APPLICATION

Table 1 demonstrates that the RC score separates the HC and Control samples, delineating highly cited authors of lower influence. We now seek other discriminators or correlates of lower influence among the HC sample using an extended set of comparative bibliometric properties obtained from the free, open and convenient bibliometric resource OpenAlex (OA, Priem et al., 2022). Our coverage comprises works published in our mathematics journal set in 2020 or later by authors in the HC sample and the Control sample. We exclude five journals (PLOS One, Science Reports, Heliyon, PNAS and PeerJ) due to many non-Mathematics works. The OA source list does not cover around 20 ESI journals. We omit two sample authors whose OA data erroneously combines several people. The citation counts capture references to works published in 2020, by works published in 2020 or later: a small number (<10) of HC and Controls have no listed OA mathematics publications in this interval.

OpenAlex data supports further exploration of RC scores. Fig. 3 exhibits the distribution of citation counts and RC scores for over 98,000 authors having at least one citation. The Controls display a wide range of citation counts, and relatively high RC scores. Around 10% of the HC authors display similar characteristics. Most HC authors, however, have RC scores scattered towards the bottom of the range, especially for authors having more than 50 citations.

Figure 3. Scatter plot (log-log) of citation counts against RC scores for the HC and Control samples, backgrounded against around 98,000 OpenAlex authors. Authors have received at least one citation in 2020 from articles published in 2020 or later in our journal set.

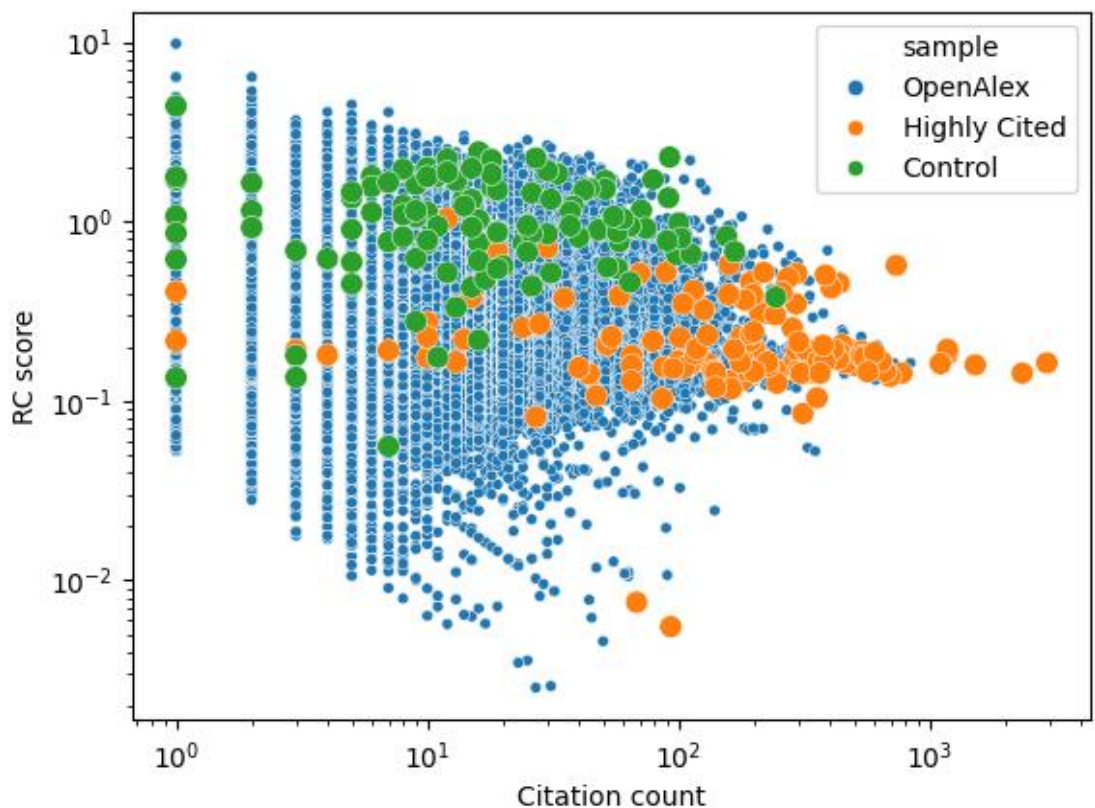

Table 2 exhibits basic information for the two samples alongside the full OpenAlex set. The HC works total refers to all works with at least one HC author, and similarity for the other totals. Note the extensive reach of the HC sample, where the HC authors produce over 7,000 works with 7,500

Page | 11

coauthors. The authors of the Control sample produce fewer than one-third of this volume. Compared with the OpenAlex community (of whom fewer than 12% averaged more than one article per year), the HC community publishes on average around fifty times more works per year, each attracting around two times more citations per year, thus accumulating 150 times more citations per work per year. While the work of only a few authors dominates these trends, ESI mathematics HCs are all highly cited and highly productive, and their high productivity ensures dominance of the top centile of the citation distribution. By contrast, relative to the OpenAlex community the Control sample publishes around twelve times more works per year, each attracting around the same number of citations per year. HC works have a broader range of institutions and countries while the Control sample has more pages and fewer authors. Kolmogorov-Smirnov tests confirm that both the HC and Control distributions underlying these means are highly disparate samples from the OpenAlex distribution.

Table 2. Summary of OpenAlex data for the journal set. A work is Highly Cited if at least one author is in the HC sample, and similarly for Control works. Totals refer to counts of distinct articles. Means are the population means of the per-article counts of unique items.

|  | Highly Cited | Control | OpenAlex |
|---|---|---|---|
| **Total** | | | |
| Articles | 7,037 | 1,944 | 373,409 |
| Journals | 428 | 364 | 922 |
| Coauthors | 7,512 | 2,301 | 421,135 |
| Institutions | 2,045 | 811 | 18,300 |
| **Mean** | | | |
| Citations | 12.5 | 6.2 | 5.7 |
| Authors | 4.6 | 3.2 | 6.8 |
| Institutions | 5.6 | 4.1 | 4.2 |
| Countries | 4.0 | 2.1 | 1.8 |
| Pages | 9.7 | 26.9 | 12.3 |
| References | 31.4 | 28.1 | 31.5 |
| References/Page | 1.9 | 1.0 | 1.9 |
| APC ($US) | 2,072 | 2,817 | 2,679 |

Table 3 ranks the ten most common journals and their publishers over the two samples. The Elsevier and Springer publisher groups appear in both lists, although there are no journals in common. HC works are more concentrated into selected journals, with 41% of works in the top ten, compared with 18% of Control works.

Table 3. Most frequent journals and their publishers.

| | Highly Cited | | | Control | | |
| --- | --- | --- | --- | --- | --- | --- |
| | Journal | Publisher | Share | Journal | Publisher | Share |
| 1 | AIMS mathematics | AIMS | 6.8% | Springer mono math | Springer | 3.0% |
| 2 | Mathematics | MDPI | 5.8% | Adv Mathematics | Elsevier BV | 2.6% |
| 3 | Chaos, Solitons & Fractals | Elsevier BV | 4.9% | Arch Rat Mech Anal | Springer Sci+Bus | 2.0% |
| 4 | Adv Difference Equations | Springer Nature | 4.8% | Int Math Res Not | OUP | 1.9% |
| 5 | Math Meth App Sci | Wiley | 4.7% | Comm Math Phys | Springer Sci+Bus | 1.7% |
| 6 | Symmetry | MDPI | 4.5% | Lecture Notes in Math | Springer | 1.6% |
| 7 | Fractal and fractional | MDPI | 3.1% | J Euro Math Soc | Euro Math Soc | 1.5% |
| 8 | Fractals | World Scientific | 2.8% | J Diff Eq | Elsevier BV | 1.5% |
| 9 | Journal of function spaces | Hindawi | 2.2% | Nonlinear Analysis | Elsevier BV | 1.2% |
| 10 | Axioms | MDPI | 1.6% | Inventiones Math | Springer Sci+Bus | 1.2% |

Table 4 compares the relative frequency of authors' institutions and countries between the samples. The Chinese Medical University in Taiwan participates in the authorship of 18.5% of all HC works, yet remarkably has no Mathematics department. Taiwan published 5,115 OpenAlex works in the census period, including eight by the Control sample and 1,339 (26%) by the Highly Cited Sample. Institutions in Saudi Arabia, Turkey, and Romania are also prominent, hosting 6,650 works co-authored by a Highly Cited researcher and 28 by a Control. For these countries and thirteen others, more than 15% of all works have a Highly Cited coauthor. The profile of institutions and countries for the Control sample partly reflects our specific regional knowledge of peer-esteemed mathematics Academicians.

Table 4. Common affiliated institutions and host countries.

| | Highly Cited | | | Control | | |
| --- | --- | --- | --- | --- | --- | --- |
| | Institution | Country | Share | Institution | Country | Share |
| 1 | China Medical University | TW | 10.5% | Princeton University | US | 3.6% |
| 2 | China Medical University Hosp | TW | 7.9% | Uni Western Australia | AU | 2.8% |
| 3 | Çankaya University | TR | 4.8% | CNRS, France | FR | 2.6% |
| 4 | Institute of Space Science | RO | 3.1% | Czech Academy Sciences | CZ | 1.9% |
| 5 | King Abdulaziz University | SA | 2.3% | Autonomous Uni Madrid | ES | 1.9% |
| 6 | University of Victoria | CA | 1.6% | Sorbonne Université | FR | 1.7% |
| 7 | Azerbaijan University | AZ | 1.5% | Oklahoma State University | US | 1.7% |
| 8 | Amirkabir University of Tech | IR | 1.4% | ETH Zurich | CH | 1.6% |
| 9 | Siirt University | TR | 1.3% | Pohang Uni Sci Tech | KR | 1.6% |
| 10 | Azarbaijan Shahid Madani Uni | IR | 1.3% | Bielefeld University | DE | 1.6% |

To explore scholarly connections between sample members we constructed the co-authorship network for each sample, where vertices are authors and edges are the number of works with both vertices among the co-authors. Not all members of the samples have co-authored with other members, and both co-authorship networks comprise unconnected subcomponents. The Highly Cited network has 103 co-authoring members and 39 others. The largest connected component covers 85 co-authors, tying together 80% of co-authoring HCs and over 50% of all HC sample members. The second largest component has only four members. By contrast, for the Control sample there are only 38 co-authoring members, and the largest connected component has seven members. Perhaps the existence of a cluster of 85 Highly Cited co-authors is one basis for concerns about co-operative manipulation mentioned in the introduction, a conjecture that is reinforced by the RC scores of the cluster: $0.34 \pm 0.41$ (s.d.) compared with $1.2 \pm 0.7$ for the four members of the second High Cited component and $1.6 \pm 1.5$ for the seven members of the largest component of the Control network.

## 4. DISCUSSION

Reputable Citation strengthens the evidence that recently, many very highly cited authors in mathematics have attracted abundant citations to bodies of work of modest influence. The RC approach does not claim to be definitive, but rather screens the large, normal population of mathematics papers to isolate a far smaller population for further consideration, ideally one that contains a low proportion of false candidates. Influential and relevant work is sought out and used by many researchers, but above all, by those who also conduct influential and relevant research. Abundant citations conferred from minor works carry less significance than the few conferred by works advancing the frontiers of the discipline. Quantity matters, but more so does quality, and RC incorporates this into measures of researchers' influence and impact. Based on the work reported above we believe that RC is a promising way to address the emerging need for automated screening of citation activity to identify exceptional and influential work as well as manipulative practices.

While the calculation of RC values for a field of research should include the major journals in terms of coverage and quality, the journal set can never be complete. For example, an important and influential advance in the subdiscipline of applied statistics may stimulate citations from a broad range of exogenous journals without necessarily prompting an equally large upsurge within the field. The RC score for in-field journals will then be erroneously low, prompting the need for further work on ways to represent citations flowing between the journal set and exogenous outlets. RC should also capture a broad range of information about the relative standing of mathematics across its institutional population. The status data we have used may be biased against small, emerging, and specialized groups especially in the global South. It would be feasible for RC to develop benefit factors designed to level this bias without weakening its screening powers.

Citation manipulation undermines the utility of citation analysis in the field (e.g., Meho, 2007; Rousseau et al., 2018) including the HCR lists published by commercial companies and university researchers (Chaignon et al., 2023; Chen, 2023; Ioannidis et al., 2019; Klein & Kranke, 2023). The limitations of HCR lists are well recognized (Aksnes, 2003; Aksnes & Aagaard, 2021; Chen, 2023; Docampo & Cram, 2019; Meho, 2007, 2022). However, despite these issues the lists have been hitherto regarded as reliable and generally aligned with the perceived reputation of highly cited

Page | 14

researchers among their peers justifying their adoption as an indicator of research quality in instruments such as Academic Ranking of World Universities (Liu & Cheng, 2005).

Now a significant new problem has arisen due to the questionable reliability of current citations as a foundation for influence assessment. Especially over the past ten years, in virtually all academic disciplines, the production rate of articles and journals has increased sharply. In mathematics, there has been a sixtyfold increase in the annual production of indexed articles since 1960 and as Price (1963) foreshadowed, the growth comes with a surge of poor-quality publications. This raises barriers to discovering high-quality articles. These can be overcome to some extent by new search tools, but these concerningly often rely on citation indexes and reference lists (e.g., Tay, 2024). The surge in output tends to distort and perhaps obliterate reliable measures of impact and excellence, especially when modest papers manage to attract citations in numbers that lie in the top centile.

To cater for the rapid growth in mathematics publication, publishers have launched mega journals and thematic journals, some with thousands of articles per year. For example, the MDPI journal Mathematics (Basel) launched in 2013 with ten articles in the first volume, rising to 1,920 in volume 12 which covers the first half of 2024. As a JCR/Q1 and SJR/Q2 journal, Mathematics reports an acceptance rate of 40%, with fewer than 20 days between submission and the decision to send the article for peer review, and around three days between acceptance and publication. Efficient, automated review and publication procedures can satisfy the growth in volume, but equal diligence is also required to manage potential manipulations that can seriously compromise publishers reputations and viability.

Presently, we find dozens of authors in mathematics publishing a sizable number of articles per year, receiving scores of citations to each one, yet there is no evidence that the works are genuinely considered influential in the field. Returning to the Tahamtan and Bornmann (2022) proposal for a Social Systems Citation Theory (SSCT), these findings refer to the social system of science. The formation of the collectives of authors that maintain this activity occurs in the psychic systems of people, and the policy systems of institutions and nations. For some researchers, the satisfaction and pleasure they gain from the public and private accolades accompanying HCR status might outweigh their evaluation of the risk of discovery and disgrace due to questionable citation activities. Others, unfortunately, might be drawn into these activities as students or junior researchers, unaware that the practices carry risks. For the top administrators of some institutions, the enhancement of status and reputation through upward movement in ranking systems might outweigh the cost of querying the scholarly standing of the contributions fueling the rise, or the risk of loss of status through discovery. Many national governments use bibliometric indicators in research assessment and research resource allocation, frequently with assistance from the national peak disciplinary bodies. They willingly acknowledge apparent improvements in national research performance but rarely question whether manipulative practices have been involved and even less accept responsibility for the environment that encourages them if they occur.

## 5. PROSPECTS

Any systematic disruption of the scholarly practice of using references to acknowledge prior work or to advocate a position undermines trust in research and strengthens the call for a forensic

Page | 15

scientometrics (Mcintosh, 2024; McIntosh & Vitale, 2024; Robinson-García, 2024). Our Reputable Citation contribution emerges from research initially conducted five decades ago on social networks and citation influence advanced through PageRank and other components of the global infrastructure of research evaluation. Over this time, confidence has grown in the ability of forensic investigations of the scholarly literature to evidence misconduct such as plagiarism, fabrication, and falsification, with growing recent attention to predatory and fake journals, authorship and editorship markets, and paper mills.

Our application of Reputable Citations shows that recent mathematics literature is archiving works that may have been reviewed, published and cited for reasons that have nothing to do with mathematics research. Work to improve and extend the method could contribute to slowing this unnecessary and wasteful development.

AUTHOR CONTRIBUTIONS

COMPETING INTERESTS

The authors have no competing interests

FUNDING INFORMATION

Vicente Safón was supported by Grant PID2022-139222NB-I00, funded by MCIN/AEI/10.13039/501100011033 and by "ERDF A way of making Europe".

Domingo Docampo was supported by the Xunta de Galicia (Centro singular de investigación de Galicia accreditation 2019-2022) and the European Union (European Regional Development Fund - ERDF).

DATA AVAILABLITY

The list of journals and their RC scores is included as Supplementary Material.

REFERENCES

Abalkina, A. (2023). Challenges posed by hijacked journals in Scopus. *Journal of the Association for Information Science and Technology*. https://doi.org/10.1002/asi.24855

Aksnes, D. W. (2003). Characteristics of highly cited papers. *Research Evaluation*, *12*(3), 159-170. https://doi.org/10.3152/147154403781776645

Aksnes, D. W., & Aagaard, K. (2021). Lone Geniuses or One among Many? An Explorative Study of Contemporary Highly Cited Researchers. *Journal of Data and Information Science*, *6*(2). https://doi.org/10.2478/jdis-2021-0019

Aspesi, C., Allen, N., Crow, R., Daugherty, S., Joseph, H., McArthur, J., & Shockey, N. (2019). SPARC* Landscape Analysis: The Changing Academic Publishing Industry–Implications for Academic Institutions.

Beall, J. (2012). Predatory publishers are corrupting open access. *Nature*, *489*(7415), 179-179. https://doi.org/10.1038/489179a

Berkhin, P. (2005). A Survey on PageRank Computing. *Internet Mathematics*, *2*(1), 73-120. https://doi.org/10.1080/15427951.2005.10129098

Biagioli, M., & Lippman, A. (2020). *Gaming the metrics: Misconduct and manipulation in academic research*. MIT Press. https://doi.org/10.7551/mitpress/11087.001.0001

Bollen, J., Rodriquez, M. A., & Van De Sompel, H. (2006). Journal status. *Scientometrics*, *69*(3), 669-687. https://doi.org/10.1007/s11192-006-0176-z

Safón, V., Docampo, D., Cram, L. (2025) Screening articles by citation reputation. Quantitative Science Studies. Advance Publication. https://doi.org/10.1162/qss_a_00355

Borgman, C. L., & Furner, J. (2002). Scholarly communication and bibliometrics. *Annual review of information science and technology*, *36*(1), 2-72. https://doi.org/10.1002/aris.1440360102

Bornmann, L., & Daniel, H. D. (2008). What do citation counts measure? A review of studies on citing behavior. *Journal of documentation*, *64*(1), 45-80. https://doi.org/10.1108/00220410810844150

Brin, S., & Page, L. (1998). The anatomy of a large-scale hypertextual web search engine. *Computer networks and ISDN systems*, *30*(1-7), 107-117. https://doi.org/10.1016/S0169-7552(98)00110-X

Bush, V. (1945). Science: The endless frontier. *Transactions of the Kansas Academy of Science (1903)*, 231-264. https://doi.org/10.2307/3625196

Catanzaro, M. (2024). Citation manipulation found to be rife in math. *Science (New York, NY)*, *383*(6682), 470-470. https://doi.org/10.1126/science.ado3859

Chaignon, L., Docampo, D., & Egret, D. (2023). In search of a scientific elite: highly cited researchers (HCR) in France. *Scientometrics*, *128*(10), 5801-5827. https://doi.org/10.1007/s11192-023-04805-3

Chan, K. C., Chang, C.-H., & Chang, Y. (2013). Ranking of finance journals: Some Google Scholar citation perspectives. *Journal of Empirical Finance*, *21*, 241-250. https://doi.org/10.1016/j.jempfin.2013.02.001

Chen, X. (2023). Does cross-field influence regional and field-specific distributions of highly cited researchers? *Scientometrics*, *128*(1), 825-840. https://doi.org/10.1007/s11192-022-04584-3

Chen, Y.-L., & Chen, X.-H. (2011). An evolutionary PageRank approach for journal ranking with expert judgements. *Journal of Information Science*, *37*(3), 254-272. https://doi.org/10.1177/0165551511402421

Chughtai, G. R., Lee, J., Mehran, M., Abbasi, R., Kabir, A., & Arshad, M. (2018). Global Citation Impact rather than Citation Count. *International Journal of Advanced Computer Science and Applications*, *9*(7). https://doi.org/10.14569/ijacsa.2018.090735

Cronin, B. (2005). A hundred million acts of whimsy? *Current Science - Bangalore*, *89*(9), 1505. https://www.jstor.org/stable/i24110169

Davison, B. D. (2000). Recognizing nepotistic links on the web. In *Artificial Intelligence for Web Search: Technical Report WS-00-01* (pp. 23-28). Menlo Park, California: The AAAI Press.

Docampo, D., & Cram, L. (2019). Highly cited researchers: a moving target. *Scientometrics*, *118*(3), 1011-1025. https://doi.org/10.1007/s11192-018-2993-2

Docampo, D., & Safón, V. (2021). Journal ratings: a paper affiliation methodology. *Scientometrics*, *126*(9), 8063-8090. https://doi.org/10.1007/s11192-021-04045-3

Fister, I., Fister, I., & Perc, M. (2016). Toward the Discovery of Citation Cartels in Citation Networks. *Frontiers in Physics*, *4*. https://doi.org/10.3389/fphy.2016.00049

Fyfe, A., Coate, K., Curry, S., Lawson, S., Moxham, N., & Røstvik, C. M. (2017). Untangling academic publishing: A history of the relationship between commercial interests, academic prestige and the circulation of research. https://doi.org/10.5281/zenodo.546100

Garfield, E. (1955). Citation Indexes for Science: A New Dimension In Documentation Through Association Of Ideas. *Science*, *122*, 108-111. http://science.sciencemag.org/content/122/3159/108.long

Safón, V., Docampo, D., Cram, L. (2025) Screening articles by citation reputation. Quantitative Science Studies. Advance Publication. https://doi.org/10.1162/qss_a_00355

Garfield, E. (1979). Is citation analysis a legitimate evaluation tool? *Scientometrics*, *1*(4), 359-375. https://doi.org/10.1007/bf02019306

Garfield, E. (1998). *On The Origins of Current Contents and ISI*. Retrieved 2016 Jan 13 from http://www.garfield.library.upenn.edu/papers/origins_cc_isi.html

Garfield, E., & Welljams-Dorof, A. (1992). Of Nobel Class: A Citation Perspective on High Impact Research Authors. *Theoretical Medicine*, *13*, 117-135. https://doi.org/10.1007/BF02163625

Gilbert, G. N. (1977). Referencing as persuasion. *Social Studies of Science*, *7*(1), 113-122. https://doi.org/10.1177/030631277700700112

Gorman, M. F., & Kanet, J. J. (2005). Evaluating operations management–related journals via the author affiliation index. *Manufacturing & Service Operations Management*, *7*(1), 3-19.

Gyöngyi, Z., Berkhin, P., Garcia-Molina, H., & Pedersen, J. O. (2006). Link spam detection based on mass estimation. 32nd International Conference on Very Large Data, Seoul, Korea.

Haveliwala, T. H. (2002). Topic-sensitive pagerank. Proceedings of the 11th international conference on World Wide Web, Hawaii USA.

Hicks, D., Wouters, P., Waltman, L., de Rijcke, S., & Rafols, I. (2015). Bibliometrics: The Leiden Manifesto for research metrics. *Nature*, *520*(7548), 429-431. https://doi.org/10.1038/520429a

Ioannidis, J. P. A., Baas, J., Klavans, R., & Boyack, K. W. (2019). A standardized citation metrics author database annotated for scientific field. *PLoS biology*, *17*(8), e3000384. https://doi.org/10.1371/journal.pbio.3000384

Jarvis, R. M., & Coleman, P. (2007). Ranking law reviews by author prominence-Ten years later. *Law Libr. J.*, *99*, 573.

Jarvis, R. M., & Coleman, P. G. (1997). Ranking law reviews: An empirical analysis based on author prominence. *Ariz. L. Rev.*, *39*, 15.

Kamvar, S. D., Schlosser, M. T., & Garcia-Molina, H. (2003). The eigentrust algorithm for reputation management in p2p networks. *Proceedings of the 12th international conference on World Wide Web*, 640-651. https://doi.org/10.1145/775152.775242

Kanellos, I., Vergoulis, T., Sacharidis, D., Dalamagas, T., & Vassiliou, Y. (2019). Impact-based ranking of scientific publications: A survey and experimental evaluation. *IEEE Transactions on Knowledge and Data Engineering*, *33*(4), 1567-1584. https://doi.org/10.1109/TKDE.2019.2941206

Kapovich, I. (2011). The Dangers of the "Author Pays" Model in Mathematical Publishing. *Notices of the AMS*, *58*(9).

Klein, A.-M., & Kranke, N. (2023). Some thoughts on transparency of the data and analysis behind the Highly Cited Researchers list. *Scientometrics*, *128*(12), 6773-6780. https://doi.org/10.1007/s11192-023-04852-w

Leydesdorff, L. (1998). Theories of citation? *Scientometrics*, *43*(1), 5-25. https://doi.org/10.1007/bf02458391

Liu, N. C., & Cheng, Y. (2005). Academic ranking of world universities: Methodologies and problems. *Higher Education in Europe*, *30*(2), 127-136. https://doi.org/10.1080/03797720500260116

Safón, V., Docampo, D., Cram, L. (2025) Screening articles by citation reputation. *Quantitative Science Studies*. Advance Publication. https://doi.org/10.1162/qss_a_00355

Massucci, F. A., & Docampo, D. (2019). Measuring the academic reputation through citation networks via PageRank. *Journal of Informetrics*, *13*(1), 185-201. https://doi.org/10.1016/j.joi.2018.12.001

Mcintosh, L. D. (2024, 2024 April 2). FoSci — The Emerging Field of Forensic Scientometrics. *The Scholarly Kitchen*. https://scholarlykitchen.sspnet.org/2024/04/02/guest-post-fosci-the-emerging-field-of-forensic-scientometrics/

McIntosh, L. D., & Vitale, C. H. (2024). Forensic Scientometrics--An emerging discipline to protect the scholarly record. *arXiv preprint arXiv:2404.00478*.

Meho, L. I. (2007). The rise and rise of citation analysis. *Physics World*, *20*(1), 32-36. https://doi.org/10.1088/2058-7058/20/1/33

Meho, L. I. (2022). Gender gap among highly cited researchers, 2014–2021. *Quantitative Science Studies*, *3*(4), 1003-1023. https://doi.org/10.1162/qss_a_00218

Merton, R. K. (1973). *The sociology of science: Theoretical and empirical investigations*. University of Chicago Press.

Page, L., Brin, S., Motwani, R., & Winograd, T. (1998). *The pagerank citation ranking: Bringing order to the web* (Stanford InfoLab, Issue. http://ilpubs.stanford.edu:8090/422/

Pendlebury, D. A. (2008). White paper: Using bibliometrics in evaluating research. *Philadelphia: Thomson Reuters*.

Pinski, G., & Narin, F. (1976). Citation influence for journal aggregates of scientific publications: Theory, with application to the literature of physics. *Information Processing & Management*, *12*(5), 297-312. https://doi.org/10.1016/0306-4573(76)90048-0

Price, D. d. S. (1963). *Big science, little science*. Columbia University Press. https://doi.org/10.7312/pric91844

Priem, J., Piwowar, H., & Orr, R. (2022). OpenAlex: A fully-open index of scholarly works, authors, venues, institutions, and concepts. *arXiv preprint arXiv:2205.01833*.

Robinson-García, N. (2024). La bibliometría forense y Los Hombres de Paco. *Anuario ThinkEPI*, *18*. https://doi.org/10.3145/thinkepi.2024.e18a08

Rousseau, R., Egghe, L., & Guns, R. (2018). *Becoming metric-wise: A bibliometric guide for researchers*. Chandos Publishing is an imprint of Elsevier, Cambridge, MA, 2018. https://doi.org/10.1016/C2017-0-01828-1

Safón, V., & Docampo, D. (2023). What are you reading? From core journals to trendy journals in the Library and Information Science (LIS) field. *Scientometrics*, *128*(5), 2777-2801. https://doi.org/10.1007/s11192-023-04673-x

Salmi, J. (2013). Can Young Universities Achieve World-Class Status? *International Higher Education*, *70*, 2-3. https://doi.org/10.6017/ihe.2013.70.8701

Small, H. (1980). Co-citation context analysis and the structure of paradigms. *Journal of documentation*, *36*(3), 183-196. https://doi.org/10.1108/eb026695

Tahamtan, I., & Bornmann, L. (2019). What do citation counts measure? An updated review of studies on citations in scientific documents published between 2006 and 2018. *Scientometrics*, *121*(3), 1635-1684. https://doi.org/10.1007/s11192-019-03243-4

Safón, V., Docampo, D., Cram, L. (2025) Screening articles by citation reputation. Quantitative Science Studies. Advance Publication. https://doi.org/10.1162/qss_a_00355

Tahamtan, I., & Bornmann, L. (2022). The Social Systems Citation Theory (SSCT): A proposal to use the social systems theory for conceptualizing publications and their citations links. *El Profesional de la Informacion*. https://doi.org/10.3145/epi.2022.jul.11

Tay, A. (2024). *Academic search and discovery tools in the age of AI and large language models: An overview of the space* AI for Research Week, Singapore. https://ink.library.smu.edu.sg/ai_research_week/Programme/Programme/1/

Thomson-ISI. (2001). *ISI Launches ISIHighlyCited.com*. Retrieved 2015 June 25 from http://www.infotoday.com/IT/apr01/news7.htm

Van Raan, A. F. J. (2005). Fatal attraction: Conceptual and methodological problems in the ranking of universities by bibliometric methods. *Scientometrics*, *62*(1), 133-143. https://doi.org/10.1007/s11192-005-0008-6

Wan, X., & Liu, F. (2014). Are all literature citations equally important? Automatic citation strength estimation and its applications. *Journal of the Association for Information Science and Technology*, *65*(9), 1929-1938. https://doi.org/10.1002/asi.23083

Welljams-Dorof, A. (1997). Quantitative citation data as indicators in science evaluations: A primer on their appropriate use. In M. Frankel & J. Cave (Eds.), *Evaluating science and scientists: An East-West dialogue on research evaluation in postcommunist Europe* (pp. 202-211). Central European University Press.

Wilsdon, J., Allen, L., Belfiore, E., Campbell, P., Curry, S., Hill, S., Jones, R., Kain, R., Kerridge, S., & Thelwall, M. (2015). The metric tide. *Report of the independent review of the role of metrics in research assessment and management*. https://doi.org/10.13140/RG.2.1.4929.1363

Zhang, F. (2017). Evaluating journal impact based on weighted citations. *Scientometrics*, *113*(2), 1155-1169. https://doi.org/10.1007/s11192-017-2510-z

Zhu, X., Turney, P., Lemire, D., & Vellino, A. (2015). Measuring academic influence: Not all citations are equal. *Journal of the Association for Information Science and Technology*, *66*(2), 408-427. https://doi.org/10.1002/asi.23179

Ziman, J. (2001). *Real science: What it is, and what it means*. Cambridge University Press. https://doi.org/10.1017/CBO9780511541391